\documentclass[sigconf]{acmart}

\usepackage[font=footnotesize,skip=8pt]{caption}
\usepackage{tikz}
\usetikzlibrary{shapes}
\usetikzlibrary{calc}
\makeatletter
\newcommand\currentcoordinate{\the\tikz@lastxsaved,\the\tikz@lastysaved}
\newcommand\currentx{\the\tikz@lastxsaved}
\newcommand\currenty{\the\tikz@lastysaved}
\makeatother
\usepackage{todonotes}
\tikzstyle{rr}=[draw,rounded corners=0.2cm,minimum width=2cm]
\tikzstyle{circ}=[draw,circle,minimum size=0.4cm]

\usepackage[amssymb]{SIunits}
\usepackage{url}

\usepackage{bbm}
\usepackage{amsmath}
\usepackage{amsthm}
\usepackage{amssymb}
\usepackage{graphicx}
\usepackage{paralist}
\usepackage{caption}
\usepackage{stmaryrd}
\usepackage{mathpartir}
\usepackage{wrapfig}
\usepackage{xspace}
\usepackage{soul}

\renewcommand\paragraph[1]{\smallskip\noindent{\bf #1}}

\newtheorem*{remark}{Remark}

\usepackage{algorithm}
\usepackage[noend]{algpseudocode}
\algrenewcommand\alglinenumber[1]{\scriptsize #1:}

\algdef{SE}[DOWHILE]{Do}{doWhile}{\algorithmicdo}[1]{\algorithmicwhile\ #1}%

\newcommand{\commentm}[1]{}

\newcommand{\MathWordStyle}[1]{\mathsf{#1}\xspace}
\newcommand{\defn}{\em}

\newcommand{\LStar}{L$^*$\xspace}

\newcommand{\MM}{\mathsf{M}}
\newcommand{\IA}{\mathsf{A}}
\newcommand{\Interface}{\mathsf{I}}
\newcommand{\SInterface}{\mathsf{I}_s}
\newcommand{\tuple}[1]{\langle #1 \rangle}
\newcommand{\States}{Q}
\newcommand{\initState}{q_\iota}
\newcommand{\Inputs}{\Sigma_i}
\newcommand{\Outputs}{\Sigma_o}
\newcommand{\EInputs}{\tilde\Sigma_i}
\newcommand{\EOutputs}{\tilde\Sigma_o}
\newcommand{\Transitions}{\Delta}
\newcommand{\Transition}{\delta}

\newcommand{\OutFunc}{\MathWordStyle{Out}}
\newcommand{\stateq}{q}
\newcommand{\Symbol}{\sigma}
\newcommand{\inp}{i}
\newcommand{\out}{o}
\newcommand{\strace}{\tau_s}

\newcommand{\mquery}{\MathWordStyle{mQuery}}
\newcommand{\mOracle}{\MathWordStyle{MOracle}}
\newcommand{\eOracle}{\MathWordStyle{EOracle}}
\newcommand{\atrace}{\tau_a}
\newcommand{\itrace}{\tau_i}
\newcommand{\async}{\MathWordStyle{async}}
\newcommand{\ITraces}{\Pi_i}
\newcommand{\Traces}[1]{\MathWordStyle{Traces}(#1)}
\newcommand{\STraces}{\Pi_s}

\newcommand{\dummyOutput}{\lambda}
\newcommand{\waitForOutput}{\MathWordStyle{wait}}
\newcommand{\errorOutput}{\MathWordStyle{err}}
\newcommand{\quietOutput}{\MathWordStyle{quiet}}

\newcommand{\MMOutput}[2]{#1(#2)}

\newcommand{\Correct}{\MathWordStyle{Correct}}
\newcommand{\FillTable}{\MathWordStyle{FillTable}}
\newcommand{\BuildMM}{\MathWordStyle{BuildMM}}
\newcommand{\AnalyzeCex}{\MathWordStyle{AnalyzeCex}}
\newcommand{\cex}{\MathWordStyle{cex}}
\newcommand{\StateReps}{S_Q}
\newcommand{\StateRep}{w_i}
\newcommand{\Expts}{E}
\newcommand{\Expt}{e}
\newcommand{\Table}{T}

\newcommand{\Naturals}{\mathbb{N}}

\newcommand{\StateBound}{B_\MathWordStyle{State}}
\newcommand{\DistinguisherBound}{B_\MathWordStyle{Dist}}
\newcommand{\Representative}{R}
\newcommand{\suffix}{\MathWordStyle{suffix}}

\newcommand{\maxtime}{\MathWordStyle{t_{max}}}
\newcommand{\mintime}{\MathWordStyle{t_{min}}}

\newcommand{\textoverline}[1]{$\overline{\mbox{#1}}$}

\newcommand{\method}[1]{\texttt{#1()}}
\newcommand{\callin}[1]{\method{#1}}
\newcommand{\callback}[1]{\textoverline{\method{#1}}}
\newcommand{\callbackNoParen}[1]{\textoverline{\texttt{#1}}}

\newcommand{\tool}{{\sc DroidStar}\xspace}
\newcommand{\toolurl}{\url{https://github.com/cuplv/droidstar}\xspace}

\newcommand{\experhd}[1]{\smallskip\noindent{\bf #1.}}

\begin{document}

\copyrightyear{2018}
\acmYear{2018}
\setcopyright{acmlicensed}
\acmConference[ICSE '18]{ICSE '18: 40th International Conference on Software Engineering }{May 27-June 3, 2018}{Gothenburg, Sweden}
\acmBooktitle{ICSE '18: ICSE '18: 40th International Conference on Software Engineering , May 27-June 3, 2018, Gothenburg, Sweden}
\acmPrice{15.00}
\acmDOI{10.1145/3180155.3180232}
\acmISBN{978-1-4503-5638-1/18/05}

\title{\tool: Callback Typestates for Android Classes}

\author{Arjun Radhakrishna}
\authornote{This work was done while Arjun Radhakrishna was employed at the University of Pennsylvania.}
\affiliation{
  \institution{Microsoft}
}

\author{Nicholas V. Lewchenko}
\affiliation{
  \institution{University of Colorado Boulder}
}

\author{Shawn Meier}
\affiliation{
  \institution{University of Colorado Boulder}
}

\author{Sergio Mover}
\affiliation{
  \institution{University of Colorado Boulder}
}

\author{Krishna Chaitanya Sripada}
\affiliation{
  \institution{University of Colorado Boulder}
}

\author{Damien Zufferey}
\affiliation{
  \institution{Max Planck Institute for Software Systems}
}

\author{Bor-Yuh Evan Chang}
\affiliation{
  \institution{University of Colorado Boulder}
}

\author{Pavol {\v C}ern{\'y}}
\affiliation{
  \institution{University of Colorado Boulder}
}

\begin{abstract}

Event-driven programming frameworks, such as Android, are based on
components with asynchronous interfaces. The protocols for interacting with
these components can often be described by finite-state machines we
dub {\em callback typestates}.  Callback typestates are akin
to classical typestates, with the difference that their outputs
(callbacks) are produced asynchronously.  While
useful, these specifications are not commonly available, because
writing them is difficult and error-prone.

Our goal is to make the task of producing callback typestates
significantly easier. We present a callback typestate assistant tool,
\tool, that requires only limited user interaction to produce a
callback typestate. Our approach is based on an active learning
algorithm, \LStar. We improved the scalability of equivalence queries
(a key component of \LStar), thus making active learning tractable on
the Android system.

 We use \tool to learn callback typestates for Android
classes both for cases where one is already provided by the
 documentation, and for cases where the documentation is unclear.
The results show that \tool learns callback typestates
accurately and efficiently.
Moreover, in several cases, the synthesized callback
typestates uncovered surprising and undocumented behaviors.

\end{abstract}


\maketitle

\renewcommand{\shortauthors}{Radhakrishna et al.}

\section{Introduction}
\label{sec:introduction}

Event-driven programming frameworks interact with client code using
callins and callbacks.
Callins are framework methods that the client invokes and callbacks
are client methods that the framework invokes.
The client-framework interaction is often governed by a protocol that
can be described by a finite-state machine we call {\em callback
  typestate}. Callback typestates are akin to classical typestates~\cite{SY86}, with the key
 difference that their outputs (callbacks) are produced asynchronously.
Our goal is to make the task of producing callback typestates
significantly easier for developers.

As an example of a callback typestate, consider a typical interaction
between a client
application and the framework when the client wants to use a
particular service.
The client asks for the service to be started by invoking an
\callin{startService} callin.
After the framework receives the callin, it asynchronously starts
initializing the service.
When the service is started and ready to be used, the framework
notifies the client by invoking a \callback{onServiceStarted}
callback.
The client can then use the service.
After the client finishes using the service, it invokes a
\callin{shutdownService} callin to ask the framework to stop the
service.

\paragraph{Callback typestates.}
Callback typestates are useful in a number of ways, but they are
notoriously hard to produce.
First, callback typestates are a form of documentation.
They tell client application programmers in what order to invoke
callins and which callback to expect.
Android framework documentation for some classes already uses pictures very
similar to callback typestates (Figure~\ref{fig:MP}).
Second, callback typestates are useful in verification of
client code.
They enable checking that a client uses the framework correctly.
Third, even though we infer the callback typestates from framework
code, they can be used for certain forms of framework verification.
For instance, one can infer typestates for different versions of the
framework, and check if the interface has changed.

Callback typestates are very hard to produce manually. On one hand,
inspecting code to see in what situation a callback arrives, and what
callins are enabled after that is error-prone.
Even developers familiar with the framework often miss corner-case
behaviors.
On the other hand,
obtaining the callback typestate with manual testing is hard. One
would need to run all sequences of callins, mixed in sequence with the
callbacks they produce.  We systematize this testing approach using an
active learning algorithm.

\paragraph{Callback typestate assistant\ \tool.}
We present a tool that makes producing callback typestates
significantly easier. Our target user is a developer who wrote an
Android class that interacts asynchronously using callbacks with
client code.
\tool is a comprehensive framework for semi-automatically inferring
callback typestates.  The required user interaction
happens in multiple steps.
In the first step, the user provides code
snippets to perform local tasks, such as code for class initialization
and code for invoking each callin (similarly as in unit tests).
This is sufficient as long as certain widely applicable assumptions
hold. First, we assume that each sequence of callins produces a
sequence of callbacks deterministically
(this assumption fails when for instance a callback has a parameter
that is ignored at first by \tool but that influences the typestate).
Second, we assume that the resulting typestate is finite.
If these assumptions fail, in the following steps, \tool asks the user
for a solution to the problem.
For instance, one way to remove non-determinism is to refine one
callback into two separate logical callbacks, based on the
parameter values.
This design allows \tool to offer the user control over the final
result while requiring only limited, local, insight from the
user.
\tool is available for download at \toolurl

\paragraph{Approach.}
We present a method for inferring typestates for Android classes.
However, our method is equally applicable in other contexts.
The core algorithm is based on Angluin's \LStar
algorithm~\cite{Angluin87} adapted to Mealy
machines~\cite{ShahbazG09}.
In this algorithm, a learner tries to learn a finite-state machine ---
in our case a
callback typestate --- by asking a teacher membership and
equivalence queries.
Intuitively, a membership query asks for outputs corresponding to a
sequence of input callins, and the equivalence query asks if the learned
typestate is correct.
We note that the teacher does not need to know the solution, but only
needs to know how to answer the queries.

The key question we answer is how to implement oracles for the
membership and equivalence queries.
We show how to implement membership queries on Android classes using
black-box testing.
Our main contribution here is an efficient algorithm for implementing the
equivalence query using membership query. The insight here is that
the number of membership queries can be bounded by a function of a new
bound we call the {\em distinguisher bound}. We empirically confirmed that for
Android classes, the {\em distinguisher bound} is significantly
smaller than the state bound used in previous
work~\cite{Wmethod,WPmethod}. Given that the number of required
membership queries depends exponentially on the distinguisher bound, the novel bound is what enables our tool
to scale to Android classes.

\paragraph{Results.}
We use \tool to synthesize callback typestates for 16 Android framework
classes and classes from Android libraries.
The results show that \tool learns callback typestates accurately and
efficiently.
This is confirmed by documentation, code inspection, and manual
comparison to simple Android applications.
The running time of \tool on these benchmarks ranged between 43
seconds and 72 minutes, with only 3 benchmarks taking more than 10
minutes.
The usefulness of the distinguisher bound was also confirmed. Concretely, using
previously known bounds, learning the callback typestate for one of
our examples (MediaPlayer)
would take more than
a year, whereas with the distinguisher bound, this example takes around 72 minutes.
Furthermore, by inspecting our typestates, we uncovered corner cases
with surprising behavior that are undocumented and might even be
considered as bugs in some cases. For instance, for the commonly used AsyncTask
class, if \callin {execute}
is called after \callin {cancel} but before the \callback{onCancelled}
callback is received, it will not throw an exception but will never
cause the asynchronous task to be run.
Section~\ref{sec:eval} presents our results in more detail.

\paragraph{Contributions.}
The contributions of this paper are:
\begin{inparaenum}[(a)]
\item We introduce the notion of callback typestates and develop an approach, based on the \LStar algorithm, to infer them.
\item We show how to implement efficiently membership and equivalence oracles required by the \LStar algorithm.
\item We evaluate our approach on examples from the Android framework, and show its accuracy and effectiveness.
\end{inparaenum}

\section{Workflow and Illustrative Example}
\label{sec:example}

\begin{figure}
\centering
\includegraphics[height=8cm]{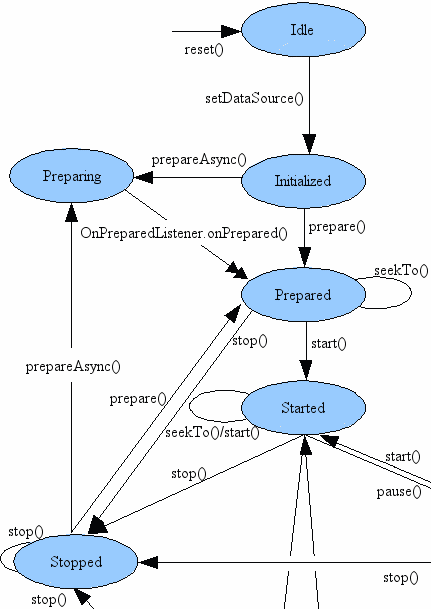}
\caption{\label{fig:MP}
  Part of MediaPlayer's callback typestate from \url{https://developer.android.com/reference/android/media/MediaPlayer.html}
}
\end{figure}

We use the Android Framework's {\tt MediaPlayer} class to explain the
standard workflow for inferring callback typestate using \tool.
This class is highly stateful---its interface includes many methods that
are only meaningful or enabled in one or two particular player
states---and makes extensive use of callins and callbacks to handle
the delays of loading and manipulating large media files.  These
properties make callback typestate a perfect fit; in fact, {\tt
MediaPlayer} has one of the very few examples
where we found a complete callback typestate
specification in the Android libraries documentation.
This callback typestate is shown in Figure~\ref{fig:MP}.

In Figure~\ref{fig:MP}, callins are
represented by single arrows and callbacks by double arrows.
Let us look at one part of the protocol that governs the
client-framework interaction.
The client first invokes the callin \callin{setDataSource}, and the
protocol transitions to the {\tt Initialized} state.
In this state, the client can invoke the callin \callin{prepareAsync},
and the protocol transitions to the {\tt Preparing} state.
In the {\tt Preparing} state, the client cannot invoke any callins, but
the framework can invoke the \callback{onPrepared} callback, and then
the protocol transitions to the {\tt Prepared} state.
At this point, the client can invoke the \callin{start} callin, and
the media starts playing.

Our goal is to semi-automatically infer the callback typestate from the
figure using the tool \tool.
The developer interacts with \tool in several steps, which we describe
now.

\subsection{Developer-Provided Snippets}
To apply \tool to the {\tt MediaPlayer} class, the developer provides a
number of code snippets detailed below that act as an interface through
which the tool can examine {\tt MediaPlayer} instances.

\paragraph{Test object and environment instantiation.}
The main callback typestate inference algorithm of \tool works roughly
by repeatedly performing tests in the form of sequences of method calls
on an object of the given class, i.e., the {\tt MediaPlayer}.
Each test must begin with an identical, isolated, class object, and if
necessary, a standard environment.
In the first step, the developer provides a snippet to
initialize such an object and environment.
In the case of {\tt MediaPlayer}, this snippet is as simple as
discarding the previous instance, creating a new one with {\tt new
MediaPlayer()}, and registering the necessary callback listeners
(explained in the {\em Callback instrumentation} paragraph below).
In some cases this snippet is more complex.  As an example, we cannot
create new instances of the {\tt BluetoothAdapter} class, so for that
class this snippet would need to bring the existing instance back to a
uniform initial state.

\paragraph{Callin declaration.} The next step is to declare the
alphabet of ``input symbols'' that represent the callins in the
interface of our class---the final callback typestate will be written
using these symbols---and map each symbol to the concrete code snippet
it represents.  In most cases, there is a one-to-one correspondence
between input symbols and callin methods.  For example,
the code snippets associated with the input symbols ${\sf \bf
  prepare}$, ${\sf \bf prepareAsync}$, and ${\sf \bf start}$ are {\tt
  prepare();}, {\tt prepareAsync();}, and {\tt start();}, respectively.

In some cases, such as when a callin takes a parameter, the developer
may instead map a symbol to a {\em set} of code snippets representing
alternative forms of the input which are suspected to have different
behavior.  In the {\tt MediaPlayer} class, the \callin {setDataSource}
callin method takes a URL argument.  The developer might (rightly)
believe that depending on the validity and reachability of the given
URL, the behavior of the callin in the typestate may differ.  In this
case, the developer may provide the two snippets {\tt
  setDataSource(goodURL);} and {\tt setDataSource(badURL);} for the
same callin.  \tool will consider both snippets for generating tests,
and further, it will indicate if they behave differently with respect
to the typestate.  In case a difference is detected, the
``non-determinism'' is handled as explained later in this section.

The complete set of input symbols which would be declared and mapped
for the {\tt MediaPlayer} class are {\sf setDataSource}, {\sf
  prepare}, {\sf prepareAsync}, {\sf start}, {\sf stop}, {\sf reset},
{\sf release}, and {\sf pause}.

\paragraph{Callback instrumentation.} As for the callin methods, which
act as the input symbols in the callback typestate, the callback methods
act as the output symbols in the callback typestate.
The developer specifies the set of output callback symbols and
associated snippets to detect when callbacks occur.
In most cases, this involved adding the listeners for the callbacks in
the initialization snippet as mentioned above. 
In the {\tt MediaPlayer} class, the output symbols are {\bf onCompleted}
and {\bf onPrepared}.

\subsection{Automated Callback-Typestate Inference}
Once the developer provides the input and output symbols
and the associated snippets, \tool attempts to automatically learn the
callback typestate following the framework of the \LStar
algorithm.

\paragraph{\LStar inference.}
In \LStar, the learner tests sequences of inputs until she can form a
consistent hypothesis automaton.
Each such test (or sequence of inputs) is called a membership query.
Once a hypothesis automaton is produced, an equivalence query is
performed; i.e., the hypothesis automaton is checked for equivalence
with the true callback typestate.
If the two are equivalent, we are done; otherwise, a counter-example
test is returned from which the tool learns.
This process repeats until the produced hypothesis automaton is correct.

For {\tt MediaPlayer}, the first set of membership queries each consist
of a single different callin.
Of these, only the query containing
\callin{setDataSource} succeeds.  The learner continues with longer
membership queries while building the hypothesis automaton.  For
instance, it learns that \callin{prepareAsync} and \callin{prepare} do
not lead to the same state: it is possible to invoke the
\callin{start} after \callin{prepare}, but not after
\callin{prepareAsync}.  Once the client receives the callback
\callback{onPrepared}, \callin{start} may be called.  The learner thus
hypothesizes a transition from the {\tt Preparing} to the {\tt
  Prepared} on \callback{onPrepared}.  Once the hypothesis is
complete, the learner asks the equivalence query.
Initially, a counter-example to equivalence is returned using which
the learner refines its hypothesis.
The final solution is found after $5$ equivalence queries.

\paragraph{Answering Equivalence Queries.}
The equivalence query, i.e., checking if a learned callback typestate is
in fact the true callback typestate is undecidable in general.
However, assuming a bound on the size of the typestate, the equivalence
query can be implemented using further testing.
However, equivalence queries are still expensive and to make them
practical we present an new optimization based on a \emph{distinguisher
bound}.
We can observe in Figure~\ref{fig:MP} that for any pair of states there
is a transition in one state which leads to an error in the other.
This corresponds to a distinguisher bound of 1.
Small distinguisher bounds arise because typestates are not random
automata but part of an API designed for ease of use and robustness.
Such APIs are coded defensively and are fail-fast~\cite{FailFast}, i.e.,
errors are not buffered but reported immediately.
Each state in the typestate has a specific function and an associated
set of callins and callbacks.
In automata terms, the alphabet is roughly the same size as the number
of states and each state has only a few transitions, making any two
states easy to distinguish.
In Section~\ref{sec:equiv}, we explain how to use the distinguisher
bound to implement equivalence queries and discuss why distinguisher
bounds are small in practice.

\subsection{Obstacles to Inference and Solutions}
The \LStar based callback typestate inference algorithm makes several
assumptions about the behavior of the class that do not always hold.
\tool is designed to detect these violations of assumptions and notify
the developer.
Here, we discuss two such assumptions, the exceptional situations that
arise when the assumptions are violated, and the additional developer
intervention needed to handle such cases.

\paragraph{Non-determinism.}
In input-output automata learning theory, non-determinism makes
learning impossible.
Non-determinism is the possibility of the same sequence of
input callins producing different sequences of output callbacks across
tests.
Non-determinism may be due to various
{\em controllable} and {\em non-controllable} factors.
Controllable factors include cases where behavior depends on if a file
exists, if a URL is reachable, etc.
On the other hand, non-controllable factors include random number
generators, device sensors, etc.
In practice, most of the non-determinism was controllable.

The main technique for handling non-determinism is via {\em refinement
of input or output alphabets}.
Here, a single callin or callback is split into multiple "logical"
inputs or outputs.

  (a)~Controllable non-determinism can be eliminated by incorporating
  the controlling factor into the inputs.
  For example, in the {\tt SQLiteOpenHelper} class, the behavior of the
  {\tt constructor} callin changes depending on if a file exists.
  However, after splitting the callin into two separate callins {\tt
  constructor/fileExists} and {\tt constructor/noFileExists}, the
  behavior of each of each callin becomes deterministic with respect to
  these callins.

  (b)~Another source of non-determinism is when the same callback is
  used to notify logically different events.
  For example, a class may use a generic {\tt onComplete} callback
  which is passed a status parameter that can have the values
  ``Success'' and ``Failure''.
  Based on this value, different further callins are enabled, leading to
  non-determinism.
  Here, the developer may manually refine the callback into two output
  symbols {\tt onEvent/Success} and {\tt onEvent/Failure}, and the
  behavior is deterministic with respect to these.

In summary, for controllable non-determinism, the onus is on the
developer to identify the source of the detected non-determinism and
provide a refinement of the input or output alphabet and corresponding
code snippets to control the source.
No general technique exists to handle non-controllable non-determinism,
but specific cases can be handled using techniques shown in
Section~\ref{sec:android}.

\paragraph{Non-regularity.}
Another basic assumption that \LStar based inference algorithm makes is
that the callback typestate under consideration is regular.
This assumption is commonly violated in request-response style behavior
of classes where the number of responses (output callbacks) invoked is
exactly equal to the number of requests (input callins).
Our solution to this problem is to restrict the learning to a subset of
the class behavior, such as inputs with at most one pending request
callin using a learning purpose~\cite{AV10}.
These restrictions makes the behavior regular and amenable to learning.

\section{The Callback Typestate Learning Problem}
\label{sec:problem}

We introduce formal models of interfaces, define the callback typestate
learning problem, and present an impossibility result about
learning typestates. Callback typestates have both inputs (corresponding
to callins) and outputs (corresponding to callbacks).
In automata theory, callback typestates can be seen as interface
automata.
Interface automata~\cite{AH01} are a well-studied model of automata that can
produce outputs asynchronously w.r.t. inputs.
We use the name callback typestates to emphasize that
they are a generalization of typestates as used in the programming
languages literature.

\subsection{Definitions and Problem Statement}

\paragraph{Asynchronous interfaces.}
Let $\Inputs$ and $\Outputs$ be the set of callins and
callbacks of an asynchronous interface.
We abstract away parameter and return values of callins and callbacks,
and model a behavior of the interface as a {\defn trace} $\itrace =
\Symbol_0 \ldots \Symbol_n \in (\Inputs \cup \Outputs)^*$.
The {\defn interface} $\Interface$ is given by
$\tuple{ \Inputs, \Outputs, \ITraces }$ where $\ITraces \subseteq
\{ \Inputs \cup \Outputs \}^*$ is the prefix-closed set of all feasible
traces of the interface.

\commentm{
\begin{example}
  In the android media-player example, one example of a trace is
  {\tt setDataSource()} \cdot {\tt prepareAsync()} \cdot {\callback{onPrepared}}
  \cdot {\tt start()} \cdot {\tt stop()}.
\end{example}
}

\paragraph{Interface automata.}
We use interface automata~\cite{AH01} to represent asynchronous interfaces.
An {\defn interface automaton} $\IA$ is given by
$\tuple{ \States, \initState, \Inputs, \Outputs, \Transitions_\IA }$ where:
\begin{inparaenum}[(a)]
\item $\States$ is a finite set of {\defn states},
\item $\initState \in \States$ is the {\defn initial state},
\item $\Inputs$ and $\Outputs$ are finite sets of {\defn input and
  output symbols}, and
\item $\Transitions_\IA \subseteq \States \times \{ \Inputs \cup
  \Outputs \} \times \States$ are a set of {\defn transitions}.
\end{inparaenum}
A {\defn trace} $\atrace$ of $\IA$ is given by
$ \Symbol_0 \ldots \Symbol_n$ if $\exists \stateq_0
\ldots \stateq_{n+1}:
\stateq_0 = \initState \wedge
\forall i.( \stateq_i, \Symbol_i, \stateq_{i+1}) \in
\Transitions_\IA$.
$\Traces{\IA}$ is the set of all traces of $\IA$.
\commentm{
We say that {\defn $\IA$ models the
typestate of $\Interface = \tuple{
\Inputs, \Outputs, \ITraces }$} if each trace of $\IA$ is a trace of
$\Interface$, and vice versa.
}

\paragraph{Problem statement.}
Given an interface $\Interface = \tuple{ \Inputs, \Outputs, \ITraces }$,
the {\defn callback typestate learning problem} is to learn an interface
automaton $\IA$ such that $\ITraces = \Traces{\IA}$.
We allow the learner to ask a {\defn membership oracle}
$\mOracle[\Interface]$ membership queries.
For a {\defn membership query}, the learner picks
$\mquery = \inp_0 \inp_1 \ldots \inp_n \in \Inputs^*$ and the membership
oracle $\mOracle[\Interface]$ returns either:
\begin{inparaenum}[(a)]
\item a trace $\atrace \in \ITraces$ whose sequence of callins is
  exactly $\mquery$, or
\item $\bot$ if no such trace exists.
\end{inparaenum}

\subsection{The Theory and Practice of Learning Typestates}
\label{sec:obstacles}

In general, it is impossible to learn callback typestates using only membership queries;
no finite set of membership queries fixes a unique interface
automaton.
However, callback typestates can be effectively learned given extra assumptions.
We now analyze the causes behind the impossibility and highlight the
assumptions necessary to overcome it.

\paragraph{Unbounded asynchrony.}
Membership queries alone do not tell us if the interface will emit more outputs (callbacks) at any point in time.
Hence, we assume:\\
\centerline{\bf Assumption 1: Quiescence is observable.}
This assumption is commonly used in ioco-testing frameworks~\cite{Vaandrager91}.
In our setting, we add an input $\waitForOutput$ and an output
$\quietOutput$, where $\quietOutput$ is returned after a
$\waitForOutput$ only if there are no other pending callbacks.
In practice, $\quietOutput$ can be implemented using
timeouts, i.e., pending callbacks are assumed to arrive
within a fixed amount of time.
If no callbacks are seen within the timeout, $\quietOutput$ is output.

\begin{example}
  Using $\waitForOutput$ and $\quietOutput$, in the MediaPlayer
  example, we have that {\tt setDataSource()} \cdot {\tt prepareAsync()}
  \cdot {\callback{onPrepared}} \cdot {\tt wait} \cdot {\tt
  quiet} is a valid trace, but {\tt setDataSource()} \cdot {\tt
  prepareAsync()} \cdot {\tt wait} \cdot {\tt quiet} is
  not.
\end{example}

\paragraph{Behavior unboundedness.}
For any set of membership queries, let $k$ be the length of the longest query.
It is not possible to find out if the interface exhibits
different behavior for queries much longer than $k$.
This is a theoretical limitation, but is not a problem in practice~\cite{10.1007/978-3-642-22655-7_2}; most callback typestates are rather small ($\leq 10$ states).\\
\centerline{\bf Assumption 2: An upper bound on the size of }
\centerline{\bf the typestate being learned is known.}

\paragraph{Non-determinism.}
We need to be able to observe the systems' behaviors to learn them and non-determinism can prevent that.
Therefore, we assume:\\
\centerline{\bf Assumption 3: The interface is deterministic.}
We assume that for every trace $\atrace$ of the interface,
there is at most one output $\out \in \Outputs$ such that $\atrace \cdot \out \in \ITraces$.
In practice, the non-determinism problem is somewhat alleviated due to the nature of callback typestates (see Section~\ref{sec:android}).
See~\cite{AV10} for a detailed theoretical discussion of how non-determinism affects learnability.

\begin{example}
  Consider an interface with traces given by $(\mathtt{input} \cdot ( \mathtt{out1} \mid \mathtt{out2} ))^*$.
  All membership queries are a sequence of {\tt input}'s;
  however, it is possible that the membership oracle never returns any
  trace containing \callbackNoParen{out2}.
  In that case, no learner will be able to learn the interface exactly.
\end{example}

\section{Learning Callback Typestates \\ using \LStar}
\label{sec:algorithm}

Given {\bf Assumption 1} and {\bf Assumption 3}, we first
build a ``synchronous closure'' of an
asynchronous interface (Section~\ref{sec:async_to_sync}).
Then, we show how to learn the synchronous closure effectively given {\bf Assumption 2} (Section~\ref{sec:lstar} and~\ref{sec:equiv}).

\subsection{From Asynchronous to Synchronous Interfaces}
\label{sec:async_to_sync}

Using {\bf Assumption 1} and {\bf 3}, we build a synchronous version
of an interface in which inputs and outputs strictly alternate following \cite{AV10}.
For synchronous interfaces, we can draw learning techniques from existing work~\cite{Angluin87,AV10,HNS03,ShahbazG09}.

Define $\EInputs = \Inputs \cup \{ \waitForOutput \}$ and $\EOutputs =
\Outputs \cup \{ \quietOutput, \dummyOutput, \errorOutput \}$.
The purpose of the extra inputs and outputs is discussed below.
For any $\strace \in (\EInputs \cdot \EOutputs)^*$, we define
$\async(\strace) = \atrace \in (\Inputs \cup \Outputs)^*$ where
$\atrace$ is had from $\strace$ by erasing all occurrences of
$\waitForOutput$, $\quietOutput$, $\dummyOutput$, and $\errorOutput$.

\paragraph{Synchronous closures.}
The {\defn synchronous closure} $\SInterface$ of an asynchronous
interface $\Interface  = \tuple{ \Inputs, \Outputs, \ITraces }$ is given
by $\tuple{ \EInputs, \EOutputs, \STraces }$ where $\EInputs$ and
$\EOutputs$ are as above, and $\STraces \subseteq (\EInputs \cdot
\EOutputs)^*$ is defined as the smallest set satisfying the following:
  \vspace{-2ex}
\[
  \begin{array}{r c l}
    \epsilon \in \STraces & & \\
    \strace \in \STraces \land \async(\strace)\cdot\inp \in \ITraces & \implies & \strace \cdot \inp \cdot \dummyOutput \in \STraces \\
    \strace \in \STraces \land \async(\strace)\cdot\out \in \ITraces & \implies & \strace \cdot \waitForOutput \cdot \out \in \STraces \\
    \strace \in \STraces \land \async(\strace)\cdot\inp \not\in \ITraces  & \implies & \strace \cdot \inp \cdot \errorOutput \in \STraces \\
    \strace \in \STraces \land \out \in \Outputs \land \async(\strace)\cdot\out \not\in \ITraces & \implies & \strace \cdot \waitForOutput \cdot \quietOutput \in \STraces \\
    \strace \in \STraces \land \strace \text{ ends in } \errorOutput & \implies & \strace \cdot \inp \cdot \errorOutput \in \STraces
  \end{array}
\]
  \vspace{-1.5ex}

Informally, in $\SInterface$:
\begin{inparaenum}[(a)]
\item Each input is immediately followed by a dummy output
  $\dummyOutput$;
\item Each output is immediately preceded by a wait
  input $\waitForOutput$;
\item Any call to an input disabled in $\Interface$ is immediately
  followed by an $\errorOutput$.
  Further, all outputs after an $\errorOutput$ are $\errorOutput$'s.
\item Any call to $\waitForOutput$ in a quiescent state is followed by
  $\quietOutput$.
\end{inparaenum}

Given $\mOracle[\Interface]$ and {\bf Assumption 1}, it is easy to
construct the membership $\mOracle[\SInterface]$.
Note that due to {\bf Assumption 3}, there is exactly one possible reply
$\mOracle[\SInterface](\mquery)$ for each query $\mquery$.
Further, by the construction of the synchronous closure, the inputs
and outputs in $\mOracle[\SInterface](\mquery)$ alternate.

\paragraph{Mealy machines.}
We model synchronous interfaces using the simpler formalism of Mealy
machines rather than interface automata.
A {\defn Mealy machine} $\MM$ is a tuple  $\tuple{ \States,
\initState, \EInputs, \EOutputs, \Transition, \OutFunc }$ where:
\begin{inparaenum}[(a)]
\item $\States$, $\initState$, $\EInputs$, and $\EOutputs$ are states,
  initial state, inputs and outputs, respectively,
\item $\Transition : \States \times \EInputs \to \States$ is a {\defn
  transition function}, and
\item $\OutFunc : \States \times \EInputs \to \EOutputs$ is an {\defn
  output function}.
\end{inparaenum}
We abuse notation and write $\OutFunc(\stateq, \inp_0\ldots\inp_n) =
\out_1\ldots\out_n$ and $\Transition(\stateq, \inp_0\ldots\inp_n) =
\stateq'$  if $\exists \stateq_0,\ldots, \stateq_{n+1} : \stateq_0
= \stateq \wedge \stateq_{n+1} = \stateq' \wedge \forall 0 \leq i \leq n :
\Transition(\stateq_i, \inp_i) = \stateq_{i+1} \wedge \OutFunc(\stateq_i,
\inp_i) = \out_i$.
A sequence $\inp_0\out_0\ldots \inp_n\out_n \in ( \EInputs \cdot
\EOutputs)^*$ is a {\defn trace} of $\MM$ if $\OutFunc(\initState,
\inp_0\ldots\inp_n) = \out_0\ldots\out_n$.
We often abuse notation and write $\MM(\inp_0\ldots\inp_n)$ instead of
$\OutFunc(\initState, \inp_0\ldots\inp_n)$.
We denote by $\Traces{\MM}$ the set of all traces of $\MM$.

\subsection{\LStar: Learning Mealy Machines}
\label{sec:lstar}

For the sake of completeness, we describe the classical \LStar learning
algorithm by Angluin~\cite{Angluin87} as adapted to Mealy machines
in~\cite{ShahbazG09}.
A reader familiar with the literature on inference of finite-state
machines may safely skip this subsection.

Fix an asynchronous interface $\Interface$ and its synchronous closure
$\SInterface$.
In \LStar, in addition to a membership oracle
$\mOracle[\SInterface]$, the learner has access to an {\defn equivalence
oracle} $\eOracle[\SInterface]$.
For an equivalence query, the learner passes a Mealy machine $\MM$ to
$\eOracle[\SInterface]$, and is in turn returned:
\begin{inparaenum}[(a)]
\item A {\defn counterexample input} $\cex = \inp_0\ldots \inp_n$
  such that $\MMOutput{\MM}{\cex} = \out_0 \ldots \out_n$ and
  $\mOracle[\SInterface](\cex) \neq \inp_0\out_0 \ldots \inp_n\out_n$, or
\item $\Correct$ if no such $\cex$ exists.
\end{inparaenum}

The full \LStar algorithm is in Algorithm~\ref{algo:lstar}.
In Algorithm~\ref{algo:lstar}, the learner maintains:
\begin{inparaenum}[(a)]
\item a set $\StateReps \subseteq \EInputs^*$ of {\defn
  state-representatives} (initially set to $\{ \epsilon \}$),
\item a set $\Expts \subseteq \EInputs^*$ of {\defn experiments}
  (initially set to $\EInputs$), and
\item an {\defn observation table} $\Table : (\StateReps \cup
  \StateReps \cdot \EInputs) \to (\Expts \to \EOutputs^*)$.
\end{inparaenum}
The observation table maps each prefix  $\StateRep$
and suffix $\Expt$ to
$\Table(\StateRep)(\Expt)$, where $\Table(\StateRep)(\Expt)$ is the
suffix of the output sequence of $\mOracle(\StateRep \cdot \Expt)$ of
length $\vert \Expt \vert$.
The entries are computed by the sub-procedure $\FillTable$.

Intuitively, $\StateReps$ represent Myhill-Nerode equivalence
classes of the Mealy machine the
learner is constructing, and $\Expts$ distinguish between
the different classes.
For $\StateReps$ to form valid set of Myhill-Nerode classes, each
state representative extended with an input, should be equivalent to
some state representative.
Hence, the algorithm checks if each $\StateRep \cdot \inp \in
\StateReps \cdot \EInputs$ is equivalent to some $\StateRep'
\in \StateReps$ (line~\ref{line:check_closed}) under $\Expts$,
and if not, adds $\StateRep \cdot \inp$ to $\StateReps$.
If no such $\StateRep \cdot \inp$ exists, the learner constructs a Mealy
machine $\MM$ using the Myhill-Nerode equivalence classes, and queries
the equivalence oracle (line \ref{line:eq_check}).
If the equivalence oracle returns a counterexample, the
learner adds a suffix of the counterexample to $\Expts$;
otherwise, it returns $\MM$.
For the full description of the choice of suffix,
see~\cite{RivestS93,ShahbazG09}.
\begin{algorithm}[t]
  \begin{algorithmic}[1]
    \Require Membership oracle $\mOracle$, Equivalence oracle
    $\eOracle$
    \Ensure Mealy machine $\MM$
    \State
        $\StateReps \gets \{ \epsilon \}$;
        $\Expts \gets \EInputs$;
        $\Table \gets \FillTable(\StateReps, \EInputs, \Expts,
        \Table)$
    \While { $\MathWordStyle{True}$ }
    \While { $ \exists \StateRep \in \StateReps, \inp \in \EInputs: \not
    \exists \StateRep' \in \StateReps: \Table(\StateRep \cdot \inp) =
    \Table(\StateRep')$ }\label{line:check_closed}
    \State
        $\StateReps \gets \StateReps \cup \{ w_i \cdot \inp \}$;
        $\FillTable(\StateReps, \EInputs, \Expts, \Table)$
    \EndWhile
    \State
        $\MM \gets \BuildMM(\StateReps, \EInputs, \Table)$;
        $\cex \gets \eOracle(\MM)$
        \label{line:eq_check}
    \If { $\cex = \Correct$ }
    \Return $\MM$
    \EndIf
    \State
        $\Expts \gets \Expts \cup \AnalyzeCex(\cex, \MM)$;
        $\FillTable(\StateReps, \EInputs, \Expts, \Table)$
    \EndWhile

    \Function {$\BuildMM$} {$\StateReps$,$\EInputs$,$\EOutputs$,$\Table$}
    \State $
        \States \gets \{ [ \StateRep ] \mid \StateRep \in \StateReps \} ;\quad
        \initState \gets [ \epsilon ]$
    \State $\forall \StateRep, \inp:
        \Transition([\StateRep], \inp) \gets [\StateRep']$~~~if
      $\Table(\StateRep \cdot \inp) = \Table(\StateRep')$
    \State $\forall \StateRep, \inp:
        \OutFunc([\StateRep], \inp) \gets \out$~~~if
      $\Table(\StateRep)(\inp) = \out$
    \State
      \Return $\tuple{ \States, \initState, \EInputs, \EOutputs,
      \Transition, \OutFunc }$
    \EndFunction
    \Function {$\AnalyzeCex$}{$\MM$,$\cex$}
    \ForAll { $0 \leq i \leq \vert \cex \vert$ and $w_i^p, w_i^s$ such
    that $w_i^p \cdot w_i^s = \cex \wedge \vert w_i^p \vert = 1$  }
    \State $w_o^p \gets \MMOutput{\MM}{w_i^p}; [ w_i^{p'} ] \gets \Transition([\epsilon], w_o^p)$
    \State $w_o^s \gets \mbox{last $\vert w_i^s \vert$ of $\OutFunc(\mOracle(w_i^{p'} \cdot w_i^s))$}$
    \If { $w_o^p \cdot w_o^s \neq \OutFunc(\mOracle(\cex))$ } \Return { $w_i^s$ } \EndIf
    \EndFor
    \EndFunction
    \Procedure {$\FillTable$} {$\StateReps$,$\EInputs$,$\Expts$,$\Table$}
    \ForAll {$w_i \in \StateReps \cup \StateReps \cdot \EInputs$, $\Expt \in \Expts$}
    \State $\Table(w_i)(\Expt) \gets \mbox{Suffix of $\OutFunc(\mOracle(w_i \cdot
    \Expt))$ of length $\vert \Expt \vert$}$
    \EndFor
    \EndProcedure
  \end{algorithmic}
  \caption{\LStar for Mealy machines}
  \label{algo:lstar}
\end{algorithm}

\begin{theorem}[\cite{ShahbazG09}]
  \label{thm:lstar}
  Let there exist a Mealy machine $\MM$ with $n$ states such that
  $\Traces{\MM}$ is the set of traces of $\SInterface$.
  Then, given $\mOracle[\SInterface]$ and $\eOracle[\SInterface]$,
  Algorithm~\ref{algo:lstar} returns $\MM$ making at most $\vert
  \EInputs \vert^2 n + \vert \EInputs \vert n^2 m$ membership and $n$
  equivalence queries, where $m$ is the maximum length of
  counterexamples returned by $\eOracle[\SInterface]$.
  If $\eOracle[\SInterface]$ returns minimal counterexamples, $m \leq
  O(n)$.
\end{theorem}

\subsection{An Equivalence Oracle Using Membership Queries}
\label{sec:equiv}

Given a black-box interface in practice, it is not feasible to directly
implement the equivalence oracle required for the \LStar algorithm.
Here, we demonstrate a method of implementing an equivalence oracle
using the membership oracle using the boundedness assumption ({\bf
Assumption 2}).
As before fix an asynchronous interface $\Interface$ and its synchronous
closure $\SInterface$.
Further, fix a target minimal Mealy machine $\MM^*$ such that $\Traces{\MM^*}$
is the set of traces of $\SInterface$.

\paragraph{State bounds.}
A {\defn state bound} of $\StateBound$ implies that the
target Mealy machine $\MM^*$ has at most $\StateBound$ states.
Given a state bound, we can replace an equivalence check with a number
of membership queries using the following theorem.
\begin{theorem}
  \label{thm:equiv_bound}
  Let $\MM$ and $\MM'$ be Mealy machines having $k$ and $k'$ states,
  respectively, such that $\exists w_i \in
  \EInputs^* : \MMOutput{\MM}{w_i} \neq \MMOutput{\MM'}{w_i'}$.
  Then, there exists an input word $w_i'$ of length at most $k + k' - 1$ such
  that $\MMOutput{\MM}{w_i'} \neq \MMOutput{\MM'}{w_i'}$.
\end{theorem}
The proof is similar to the proof of the bound $k + k' - 2$ for finite
automata (see \cite[Theorem 3.10.5]{shallit08}).
We can check equivalence of $\MM^*$ and any given $\MM$
by testing that they have equal outputs on all inputs of length
at most $k_\MM + \StateBound - 1$, i.e., using $O(\vert \EInputs
\vert^{\StateBound + k - 1})$ membership queries.
\commentm{
\begin{theorem}
  Fix a state bound of $\StateBound$ for the target Mealy machine
  $\MM^*$.
  Given a membership oracle $\mOracle[\SInterface]$ and a Mealy machine
  $\MM$ with $k$ states, the equivalence query can be answered using at
  most $\vert \EInputs \vert^{\StateBound + k - 1}$ membership
  queries.
\end{theorem}
}
While this simple algorithm is easy to implement, it is
inefficient and the number of required membership queries make it
infeasible to implement in practice.
Other algorithms based on state bounds have a similar problems with
efficiency (see Remark in Section~\ref{sec:equiv}).
Further, the algorithm does not take advantage of the structure of
$\MM$.
The following discussion and algorithm rectifies these short-comings.

\paragraph{Distinguisher bounds.}
A {\defn distinguisher bound} of $\DistinguisherBound \in \Naturals$
implies that for each pair of states $\stateq_1^*, \stateq_2^*$ in the
target Mealy machine $\MM^*$ can be distinguished by an input word $w_i$
of length at most $\DistinguisherBound$, i.e.,
$\OutFunc^*(\stateq_1^*, w_i) \neq \OutFunc^*(\stateq_2^*, w_i)$.
Intuitively, a small distinguisher bound implies that each state
is ``locally'' different, i.e., can be
distinguished from others using small length input sequences.
The following theorem shows that a state bound implies a comparable
distinguisher bound.
\begin{theorem}
  \label{thm:state_to_dist}
    State bound $k$ implies distinguisher bound $k - 1$.
\end{theorem}

\paragraph{Small distinguisher bound.}
In practice, distinguishers are much smaller than the bound implied by the state bound.
For the media-player, the number of states is $10$, but only distinguishers of length $1$ are required.
This pattern tends to hold in general due to the following principles of good interface design:
\begin{compactitem}
  \item \emph{Clear separation of the interface functions.}
        Each state in the interface has a specific function and a specific set of callins and callbacks.
        There is little reuse of names across state.
        The typestate's alphabet is roughly the same size as the number of states.
  \item \emph{Fail-fast}.
        Incorrect usage of the interface is not silently ignored but reported as soon as possible.
        This makes it easier to distinguish states as disabled callins
        lead directly to errors.
  \item \emph{No buffering}.
        More than just fail-fast, a good interface is interactive and the effect of callins must be immediately visible rather than hidden.
        A good interface is not a combination lock that requires many
        inputs that are silently stored and only acknowledged at the very end.
\end{compactitem}
This observation also is not specific to callbacks typestates and it has been already observed for libraries~\cite{Caso:2013:EPA:2491509.2491519}. 

\paragraph{Equivalence algorithm.}
Algorithm~\ref{algo:equiv_dist_bound} is an equivalence oracle for Mealy
machines using the membership oracle, given a distinguisher bound.
First, it computes state representatives
$\Representative : \States \to \EInputs^*$:
for each $\stateq \in \States$, $\Transition(\initState,
\Representative(\stateq)) = \stateq$ (line~\ref{line:compute_reps}).
Then, for each transition in $\MM$, the algorithm first checks whether
the output symbol is correct (line~\ref{line:output_compare}).
Then, the algorithm checks the ``fidelity'' of the transition up to the
distinguisher bound, i.e., whether the representative of the previous
state followed by the transition input, and the representative of the
next state can be distinguished using a suffix of length at most
$\DistinguisherBound$.
If so, the algorithm returns a counterexample.
If no transition shows a different result, the algorithm returns $\Correct$.

Two optimizations further reduce the number of membership queries:
\begin{inparaenum}[(a)]
\item {\em Quiescence transitions.}
  Transitions with input $\waitForOutput$ and output $\quietOutput$ need
  not be checked at line~\ref{line:check}; it is a no-op at the
  interface level.
\item {\em Error transitions.}
  Similarly, transition with the output $\errorOutput$ need not be
  checked as any extension of an error trace can only have error
  outputs.
\commentm{
\item {\em Representative transitions.}
  If $\Representative(\stateq)\cdot \inp = \Representative(\stateq')$
  at line~\ref{line:check}, we once again need not check
  the transition.
  Using state representatives from \LStar instead of
  computing them at line~\ref{line:compute_reps} usually leads to a
  larger number of such cases.
}
\end{inparaenum}

\begin{remark}
  Note that if Algorithm~\ref{algo:equiv_dist_bound} is being called
  from Algorithm~\ref{algo:lstar}, the state representatives from \LStar
  can be used instead of recomputing $R$ in
  line~\ref{line:compute_reps}.
  Similarly, the counterexample analysis stage can be skipped in the
  \LStar algorithm,
  and the relevant suffix
  can be directly returned ($\suffix$ in lines~\ref{line:return1}
  and~\ref{line:return2}; and $\inp$ in line~\ref{line:output_compare}).
\end{remark}

\begin{algorithm}
  \begin{algorithmic}[1]
    \Require Mealy machine $\MM = \tuple{ \States, \initState, \EInputs,
    \EOutputs, \Transition, \OutFunc }$, Distinguisher bound
    $\DistinguisherBound$, and Membership oracle $\mOracle$
    \Ensure $\Correct$ if $\MM = \MM^*$, or $\cex \in \EInputs^*$ s.\ t.\
    $\MMOutput{\MM}{\cex} \neq \MMOutput{\MM^*}{\cex}$

    \ForAll { $\stateq \in \States$ }
    $\Representative(\stateq) \gets w_i \mid
    \Transition(\initState, w_i) = \stateq$ s.\ t.\ $|w_i|$ is minimal
    \EndFor \label{line:compute_reps}

    \ForAll { $\stateq \in \States, \inp \in \EInputs$ }
    \State
         $w_i \gets \Representative(\stateq) \cdot \inp$
    \If { $\OutFunc(\stateq, \inp) \neq \text{ last symbol of } \OutFunc(\mOracle(w_i \cdot \inp))$}
    \State
        \Return $\Representative(\stateq) \cdot \inp$
    \label{line:output_compare}
    \EndIf
    \State
        $\stateq' \gets \Transition(\stateq, \inp)$;
        $~~~~ w_i' \gets \Representative(\stateq')$
    \State $\suffix \gets \MathWordStyle{check}(w_i, w_i')$ \label{line:check}
    \If {$\suffix \neq \Correct$}
    \If {$\MMOutput{\MM}{\Representative(\stateq) \cdot \inp \cdot
    \suffix} \neq \OutFunc(\mOracle(\Representative(\stateq) \cdot \inp \cdot
    \suffix))$}
    \State \Return $\Representative(\stateq) \cdot \inp \cdot \suffix$ \label{line:return1}
    \EndIf
    \State {\bf else} \Return $\Representative(\stateq') \cdot \suffix$ \label{line:return2}
    \EndIf
    \EndFor
    \State \Return $\Correct$

    \Function {$\MathWordStyle{check}$}{$w_i$, $w_i'$}
    \ForAll { $\suffix \in \EInputs^{\leq \DistinguisherBound}$ }
    \State
         $w_o \gets \OutFunc(\mOracle(w_i \cdot \suffix))$
     \State
         $w_o' \gets \OutFunc(\mOracle(w_i' \cdot \suffix))$
    \If {the last $\vert \suffix \vert$ symbols of $w_o$ and $w_o'$ differ}
    \State
        \Return $\suffix$
    \EndIf
    \EndFor
    \State \Return $\Correct$
    \EndFunction
  \end{algorithmic}
  \caption{Equivalence oracle with distinguisher bound}
  \label{algo:equiv_dist_bound}
\end{algorithm}

\begin{theorem}
  \label{thm:algo_correctness}
  Assuming the distinguisher bound of $\DistinguisherBound$ for the
  target Mealy machine $\MM^*$, either
  \begin{inparaenum}[(a)]
  \item Algorithm~\ref{algo:equiv_dist_bound} returns $\Correct$ and
    $\forall w_i \in \EInputs^*: \MM(w_i) = \MM^*(w_i)$, or
  \item Algorithm~\ref{algo:equiv_dist_bound} returns a
    counterexample $\cex$ and $\MMOutput{\MM}{\cex} \neq
    \MMOutput{\MM^*}{\cex}$.
  \end{inparaenum}
  Further, it performs at most $\vert \States \vert \cdot \vert \EInputs
  \vert^{\DistinguisherBound + 1}$ membership queries.
\end{theorem}

\commentm{
Algorithm~\ref{algo:equiv_dist_bound} and
Theorem~\ref{thm:state_to_dist} give us improved complexity for
state bounds.
\begin{corollary}
  Assuming a state bound $\StateBound$, equivalence checking for a
  Mealy machine can be implemented using at most $\vert \States \vert
  \cdot \vert \EInputs \vert^{\StateBound}$ membership queries.
\end{corollary}
}

\begin{remark}[Relation to conformance testing algorithms]
  \label{rem:conformance}
  Note that the problem being addressed here, i.e., testing the
  equivalence of a given finite-state machine and a system whose
  behavior can be observed, is equivalent to the conformance testing
  problem from the model-based testing literature.
  However, several points make the existing conformance testing
  algorithms unsuitable in our setting.

  Popular conformance testing algorithms, like the
  W-method~\cite{Wmethod} and the W$_p$-method~\cite{WPmethod}, are based
  on state bounds and have an unavoidable $O(\vert \EInputs
  \vert^{\StateBound})$ factor in the complexity.
  In our experiments, the largest typestate had $10$ states and $7$
  inputs.
  The $O(\vert \EInputs \vert^{\StateBound})$ factor leads to an
  infeasible (i.e., $> 10^8$) number of membership queries.
  However, since distinguisher bounds are often much smaller than state
  bounds, $O(\vert \EInputs \vert^{\DistinguisherBound})$
  membership queries are feasible (i.e., $~ 10^3$).
  The W- and W$_p$-methods cannot directly use
  distinguisher bounds.

  The other common algorithm, the D-method~\cite{Hennie64,Gonenc70},
  does not apply in our setting either.
  The D-method is based on building a distinguishing sequence,
  i.e., an input sequence which produces a different sequence of
  outputs from every single state in the machine.
  However, for callback typestates, such single distinguishing
  sequences do not exist in practice.
  For similar reasons, conformance testing algorithms such as the
  UIO-method~\cite{SabnaniD85} do not apply either.

  In this light, we believe that Algorithm~\ref{algo:equiv_dist_bound}
  is a novel conformance testing algorithm useful in specific settings
  where resets are inexpensive and systems are designed to have
  small distinguisher bounds.
\end{remark}

\subsection{Putting It All Together}

We now present the full callback typestate learning solution.

\begin{theorem}
  Given a deterministic interface $\Interface$ with
  observable quiescence and the membership oracle
  $\mOracle[\Interface]$.
  Assume there exists an interface automaton
  $\IA$ with $n$ states with distinguisher bound
  $\DistinguisherBound$ modeling the typestate of $\Interface$.
  Interface automaton $\IA$ can be learned with
  $O(\vert \Inputs \vert \cdot n^3 + n \cdot \vert \Inputs
  \vert^{\DistinguisherBound})$ membership queries.
\end{theorem}

\paragraph{Proof sketch.}
Starting with an asynchronous interface $\Interface$ and a membership
oracle $\mOracle[\Interface]$, using {\bf Assumption 1} and
{\bf Assumption 3} we can construct the membership oracle
$\mOracle[\SInterface]$ for the synchronous closure $\SInterface$ of
$\Interface$.
Given the distinguisher bound (or a state bound using {\bf Assumption
2} and Theorem~\ref{thm:state_to_dist}), we can construct an equivalence
oracle $\eOracle[\SInterface]$ using
Algorithm~\ref{algo:equiv_dist_bound}.
Oracles $\mOracle[\SInterface]$ and $\eOracle[\SInterface]$ can then be
used to learn a Mealy machine $\MM$ with the same set of traces as
$\SInterface$.
This Mealy machine can be converted into the interface automata
representing the callback typestate of $\Interface$ by:
\begin{inparaenum}[(a)]
\item Deleting all transitions with output $\errorOutput$ and all
  self-loop transitions with output $\quietOutput$, and
\item Replacing all transitions with input $\waitForOutput$ with the
  output of the transition.
\end{inparaenum}

\section{Active Learning for Android}
\label{sec:android}

We implemented our method in a tool called \tool.
In this section we describe how it works, the practical challenges we faced when working with Android, and our solutions to overcome them.
\tool is implemented as an Android application and learns callback
typestates from within a live Android system.

\subsection{Designing an Experiment}

To learn a typestate, a \tool user creates a test configuration (an
extension of the {\tt LearningPurpose} class) providing necessary
information about a Java class under study.
If known, the distinguisher bound can be provided here directly;
otherwise, it can be obtained from {\bf Assumption 2} by
Theorem~\ref{thm:state_to_dist}.
The \emph{instrumented alphabet}, also defined here, specifies an
abstract alphabet for the learning algorithm and translation between
the abstract alphabet and concrete callins/callbacks of the
class under study.
Several other options are available for adjusting the learning, the
most important being the \emph{quiescence timeout} which determines
{\bf Assumption 1}.

\subsection{Observing Asynchronous Callbacks}

In our approach we assume \emph{bounded asynchrony} ({\bf Assumption 1}) and, therefore, we can
observe when the interface does not produce any new output (quiescence).
We enforce this assumption on a real system with timeouts:
the membership query algorithm waits for a new output for a fixed amount
of time $\maxtime$, assuming that quiescence is reached when this time is elapsed.
However, Android does not provide any worst case execution time for
the asynchronous operations and we rely on the user to choose a large enough $\maxtime$.
The membership query also assumes the existence of a minimum time $\mintime$ before a callback occurs.
This ensures that we can issue a membership query with two consecutive callins (so, without a $\waitForOutput$ input in between), i.e., we have the time to execute the second callin before the output of the first callin.

Consider the MediaPlayer example from
Section~\ref{sec:example}.
The membership query {\tt setDataSource(URL)} \cdot $\waitForOutput$
\cdot \callin{prepareAsync} \cdot $\waitForOutput$ may not return the
\callback{onPrepared} if $\maxtime$ is violated, i.e., if the callback
does not arrive before the timeout, and while testing it is possible that the \callin{prepareAsync} \cdot \callin{start} might not return an error as expected if the lower bound $\mintime$ is violated.
To avoid such issues we try to control the execution environment and parameters to ensure that callbacks occurred between $\mintime$ and $\maxtime$.
In the MediaPlayer case, we must pick the right media source file.

\subsection{Checking and Enforcing Our Assumptions}

The simplest experiment to learn a class's callback typestate ties a single input symbol to each of its callins and a single output symbol to each of its callbacks.
However, many Android classes have behaviors which cause this simple experiment to fail and require more detailed experiments to succeed.

The main challenges when designing an experiment are
\begin{inparaenum}[(a)]
\item \emph{Non-deterministic behaviors}, i.e., the state of the device
  and external events may influence an application. These elements are inherently
    non-deterministic; however, non-determinism violates {\bf Assumption 3}.
\item The \emph{parameter space} required to drive concrete
  test cases to witness a membership query is potentially infinite.
  Though we have ignored callin parameters till now, they are a
  crucial issue for testing.
\item The protocol we are learning may not be a regular language.
  Note that this is a violation of {\bf Assumption 2}.
\end{inparaenum}

\paragraph{Non-Deterministic Behavior.}
Non-deterministic behavior is disallowed by our {\bf Assumption 3}.
However, to make this assumption reasonable we must make non-determinism straightforward to eliminate when it arises.
We explain two primary classes of non-deterministic behaviors and strategies to eliminate these behaviors.
The first class is related to controllable inputs and the second to
uncontrollable ones (such as inputs from the device sensors).

Because the learning algorithm cannot learn from non-deterministic
systems, \tool will terminate if such behavior is detected. To assist in this process, \tool will report a non-deterministic behavior is
detected and display the disagreeing
sequences to the user.
It detects this by caching all membership queries as input/output sequence pairs.
When a new trace is explored, \tool checks that the trace prefixes are compatible with the previously seen traces.

\begin{figure}
  \centering
  \begin{tikzpicture}[node distance=.7cm,font=\small]
    \node (init1) [draw,circ] {};
    \node (created) [draw,circ,right of=init1,xshift=1cm] {};
    \node (get) [draw,circ,right of=created,xshift=0.6cm] {};
    \node (a) [right of=get,xshift=1cm, yshift=0.3cm] {};
    \node (b) [right of=get,xshift=1cm, yshift=-0.3cm] {};

    \draw[->] (init1) -- node[above] {{\tt ctor}} (created);
    \draw[->] (created) -- node[above] {{\tt getRDB}} (get);
    \draw[->] (get) -- node[above,sloped] {{\tt onCreate}} (a);
    \draw[->] (get) -- node[below,sloped] {{\tt \st{onCreate}}} (b);

    \node (init2) [draw,circ,below of=init1,yshift=-8mm] {};
    \node (created1) [draw,circ,right of=init2,xshift=1cm,yshift=0.3cm] {};
    \node (created2) [draw,circ,right of=init2,xshift=1cm,yshift=-0.3cm] {};
    \node (get1) [draw,circ,right of=created1,xshift=0.6cm] {};
    \node (get2) [draw,circ,right of=created2,xshift=0.6cm] {};
    \node (c) [right of=get1,xshift=1cm] {};
    \node (d) [right of=get2,xshift=1cm] {};

    \draw[->] (init2) -- node[above,sloped] {{\tt ctor/fe}} (created1.west);
    \draw[->] (init2) -- node[below,sloped] {{\tt ctor/nfe}} (created2.west);
    \draw[->] (created1) -- node[above] {{\tt getRDB}} (get1);
    \draw[->] (created2) -- node[above] {{\tt getRDB}} (get2);
    \draw[->] (get1) -- node[above,sloped] {{\tt onCreate}} (c);
    \draw[->] (get2) -- node[above,sloped] {{\tt \st{onCreate}}} (d);
  \end{tikzpicture}
  \caption{Eliminating non-determinism in {\tt SQLiteOpenHelper}}
  \label{fig:sqlite}
\end{figure}
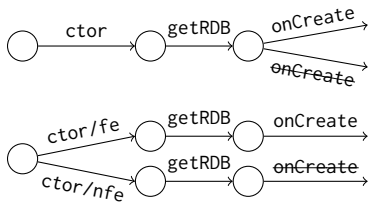

In the first case, a hidden (not modeled) controllable input
influences the typestate.
We resolve this non-determinism by manually adding the input value and
create a finer alphabet that explicate the previously hidden state of
the environment.
For example, in the class {\tt SQLiteOpenHelper}, the {\tt
getReadableDatabase()} may either trigger a \callback{onCreate} callback or
not, depending on the parameter value to a previous callin ({\tt
constructor})was the name of an existing database file.
Hence, the behavior of the callin is non-deterministic, depending on the
status of the database on disk.
In the {\tt SQLiteOpenHelper} example, we split the {\tt
constructor} callin into {\tt constructor/fileExists} and {\tt
constructor/noFileExists} and pass the right parameter values in each
case.
With this extra modeling we can learn the interface automaton, since the
execution {\tt getReadableDatabase()} ends in two different states of the
automaton (see Figure~\ref{fig:sqlite}).

The second class is the effect of the uncontrollable inputs on a
typestate.
Such effects, by definition, cannot be controlled or made explicit prior to the call.
We can sometimes to remove this non-determinism by merging different
outputs, considering them to be the same.
This is the dual of the previous solution.

An example is the {\tt SpeechRecognizer}, for which calling \callin{startListening} produces different callbacks depending on the environment.
As the environment cannot be reasonably controlled, we merge outputs to go to the same state.
If outputs are erroneously merged, the non-determinism will propagate and continue to manifest.
Thus there is no risk of unsound results. 

\paragraph{Handling Callin Parameters.}
While parameter-less callins such as \callin{start} and \callin{stop} are common in Android classes, many parameterized callins exist.
Because input symbols need to be listed in the experiment definition, the full range of parameter values cannot be explored.
In practice, we found that parameters often have little effect on the typestate automaton.
In cases where they do affect the automaton, multiple input symbols can be defined to represent the same method called with several different parameters.
This solution is similar to splitting on environmental effects when dealing with non-determinism.

\paragraph{Learning from Non-Regular Languages.}
An intrinsic limitation of \LStar is that it learns only regular languages.
However, some classes expose non-regular protocols.
Common cases include situations where a \emph{request} callin invoked
$n$ times trigger exactly $n$ {\em response} callbacks.
In the SpellCheckerSession class, callin \callin{getSuggestion} and callback \callback{onGetSuggestions} follow this pattern.

However, even in such cases, it can be useful to build a regular approximation of the typestate.
For example, restricting the typestate to behaviors where there is at most one pending request (a regular subset) provides all the information a programmer would need.
Hence, in such cases, we use the technique of \emph{learning purposes}~\cite{AV10} to learn a regular approximations of the infinite typestate.

\section{Empirical Evaluation}
\label{sec:eval}

We evaluated our interface-learning technique, as implemented in
\tool, by using it to generate callback typestates for $16$ classes,
sampled from the Android Framework and popular third-party libraries.
\tool is available at \toolurl.
For these experiments, \tool was run on an LG Nexus~5 with Android
framework version 23.  Our evaluation was designed to answer the
following questions:

\begin{compactenum}
\item Does our technique learn typestates efficiently?
\item What size distinguisher bounds occur in practice?  Do they
  support the small distinguisher bound hypothesis?
\item Do the callback typestates we learn reveal interesting or
  unintended behavior in the interfaces?
\end{compactenum}

\paragraph{Methodology.}
For each experimental class, we manually identified a reduced alphabet
of relevant callins and callbacks and provided them (along with other
necessary information as explained in Section~\ref{sec:android}) to
\tool through instances of the {\tt LearningPurpose}.
Relevant callins and callbacks for these experiments were those which,
according to the available documentation, appeared to trigger or
depend on typestate changes (enabling or disabling of parts of the
interface).
Each instance consisted of $50-200$ lines of, mostly boiler-plate,
Java or Scala code.

To evaluate efficiency, we measured the overall time taken for
learning, as well as the number of membership (MQ) and equivalence
queries (EQ).  The number of queries is likely a better measure of
performance than running time: the running time depends on external
factors.  For example, in the media player the running time depends on
play-length of the media file chosen during testing.

We validated the accuracy of learned callback typestates using two
approaches.  First, for classes whose documentation contains a picture
or a description of what effectively is an callback typestate, we
compared our result to the documentation.  Second, for all other
classes we performed manual code inspection and ran test apps to
evaluate correctness of the produced typestates.

We used a distinguisher bound of $2$ for our experiments; further, we
manually examined the learned typestate and recorded the actual
distinguisher bound.  For our third question, i.e., does the learned
callback typestate reveal interesting behaviors, we manually examined
the learned typestate, compared it against the official Android
documentation, and recorded discrepancies.

\begin{table*}[t]
  \centering
\begin{tabular}[pos]{|l|r|r|r|r|r|r|r|}
\hline
  Class name           & LP LoC   & states  & Time (s) & MQ            & EQ & MQ per EQ  & $\DistinguisherBound$ (needed)  \\
\hline
AsyncTask              &  79      & 5       &   49     &   372   (94)  & 1  &  356   (0) & 2 (1) \\
BluetoothAdapter       & 161      & 12      & 1273     &   839  (157)  & 2  &  420  (16) & 2 (1) \\
CountDownTimer         &  94      & 3       &  134     &   232   (61)  & 1  &  224   (0) & 2 (1) \\
DownloadManager        &  84      & 4       &  136     &   192   (43)  & 1  &  190   (0) & 2 (1) \\
FileObserver           & 134      & 6       &  104     &   743  (189)  & 2  &  351   (8) & 2 (1) \\
ImageLoader (UIL)      &  80      & 5       &   88     &   663  (113)  & 2  &  650  (33) & 2 (1) \\
MediaCodec             & 152      & 8       &  371     &  1354  (871)  & 1  &  973 (482) & 2 (1) \\
MediaPlayer            & 171      & 10      & 4262     & 13553 (2372)  & 5  & 2545 (384) & 2 (1) \\
MediaRecorder          & 131      & 8       &  248     &  1512  (721)  & 1  & 1280 (545) & 2 (1) \\
MediaScannerConnection &  72      & 4       &  200     &   403  (161)  & 2  &  163  (57) & 2 (1) \\
OkHttpCall (OkHttp)    &  79      & 6       &  463     &   839  (166)  & 2  &  812  (13) & 2 (2) \\
RequestQueue (Volley)  &  79      & 4       &  420     &   475  (117)  & 1  &  460   (0) & 2 (1) \\
SpeechRecognizer       & 168      & 7       & 3460     &  1968  (293)  & 3  &  646  (35) & 2 (1) \\
SpellCheckerSession    & 109      & 6       &  133     &   798  (213)  & 4  &  374   (8) & 2 (1) \\
SQLiteOpenHelper       & 140      & 8       &   43     &  1364  (228)  & 2  &  665   (6) & 2 (2) \\
VelocityTracker        &  63      & 2       &   98     &  1204  (403)  & 1  & 1156   (0) & 2 (1) \\
\hline
\end{tabular}
\caption{\tool experimental results.}
\label{table:results}
  \vspace{-2ex}
\end{table*}

\subsection{Results}
We discuss the results (in Table~\ref{table:results}) and
our three questions.

\noindent{\em Question~$1$: Efficiency.}  The table shows that our
technique is reasonably fast: most typestates learned within a few
minutes.  The longest one takes $71$ minutes, still applicable to
nightly testing.  The numbers for membership queries are reported as
$X (Y)$---$X$ is the number of membership queries asked by the
algorithm, while $Y$ is the number actually executed by the membership
oracle.  This number is lower as the same query may be asked multiple
times, but is executed only once and the result is cached.
For each benchmark, the accuracy validation showed that the
produced typestate matched the actual behavior.

\noindent{\em Question~$2$: Distinguisher Bounds.}  As mentioned
before, we used a distinguisher bound of $2$ for all experiments.
However, a manual examination of the learned callback typestates
showed that a bound of $1$ would be sufficient in all
cases except the {\tt SQLiteOpenHelper} and the {\tt OkHttpCall}
where bounds of $2$ are necessary.  This supports our conjecture that,
in practice, interfaces are designed with each state having a unique
functionality (see Section~\ref{sec:equiv}).

\noindent{\em Question~$3$: Interesting Learned Behavior.}
Of the three questions, our experiments to examine the learned
callback typestate for interesting behavior turned out to be the
most fruitful, uncovering several discrepancies, including corner cases,
unintended behavior and likely bugs, in the Android framework.
These results reaffirm the utility of our main goal of automatically
learning callback typestate, and suggest that learning typestate can
serve valuable roles in documentation and validation of callback
interfaces.

In $2$ cases, the learned typestate and documented behavior differed in
certain corner cases.
We carefully examined the differences, by framework
source examination and manually writing test applications, and found
that the learned typestate was correct and the documentation
was faulty.
In $5$ other cases, we believe the implemented behavior is not the
intended behavior, i.e., these are likely bugs in the Android
implementation.
These discrepancies mostly fall into two separate categories:

\noindent{\em Incorrect documentation.}
In such cases, it turned out that the discrepancy is minor and unlikely to
produce bugs in client programs.

\noindent{\em Race conditions.}
Several likely bugs were due to a specific category of race conditions.
These interfaces have
\begin{inparaenum}[(a)]
\item a callin to start an action and a corresponding callback which
  is invoked when the action is successfully completed;
\item a callin to cancel an already started action and a corresponding
  callback which is invoked if the action is successfully cancelled.
\end{inparaenum}
When the start action and cancel action callins are called in sequence, the
expectation is that exactly one of the two callbacks are called.
However, when the time between the two callins is small, we were able to
observe unexpected behaviors, including neither or both callbacks being
invoked.

\subsection{Selected Experiments}
Of our $16$ benchmarks, we briefly explain $5$ here.  The remaining experiments are discussed in the technical report~\cite{DBLP:journals/corr/RadhakrishnaLMM17}\footnote{\url{http://arxiv.org/abs/1701.07842}}.

\experhd{MediaPlayer}
This is the class from the example in
Section~\ref{sec:example}.
The learned typestate differs from the existing documentation.
The learned typestate:
\begin{inparaenum}[(a)]
\item has the \callin{pause} callin enabled in
  the ``playback completed'' state, and
\item shows that \callback{onPrepared} is
  invoked even after the synchronous callin \callin{prepare}.
\end{inparaenum}
Though undocumented, these behaviors are unlikely to cause any issues.

\experhd{AsyncTask}
The AsyncTask class turns arbitrary computations into callback operations
with progress tracking and results are delivered via callbacks.  For our
experiment, the computation is a simple timer.  A constructed AsyncTask object
performs its task when it receives the \callin {execute} callin, and then
either returns the results
with the
\callback{onPostExecute} callback, or returns an \callback{onCancelled} if
\callin {cancel} is called first.  The object is single-use; after it has
returned a callback it will accept no further \callin {execute} commands.

\begin{figure}
  \begin{tikzpicture}[node distance=1.8cm,font=\small]
    \node (0) [rr] {Start};
    \node (1) [rr,right of=0,xshift=1cm] {Cancelling};
    \node (2) [rr,below of=0,yshift=.5cm] {Running};
    \node (3) [rr,right of=2,xshift=1cm] {Cancelling'};
    \node (4) [rr,right of=3,xshift=1.2cm] {Completed};

    \draw[->]
            (-2cm, 0 |- 0)
            -- (0.west);
    \path[->] (0) edge[          ] node[left] {{\tt execute()}} (2) ;
    \path[->] (1) edge[          ] node[left] {{\tt execute()}} (3) ;
    \path[->] (0) edge[          ] node[above,sloped,yshift=.2cm] {{\tt cancel()}} (1) ;
    \path[->] (2) edge[          ] node[below,sloped,yshift=-.2cm] {{\tt cancel()}} (3) ;
    \path[->] (3) edge[          ] node[below,sloped,yshift=-.2cm] {\textoverline{{\tt onCancelled()}}} (4) ;
    \path[->] (1) edge[          ] node[right,yshift=.1cm,xshift=.1cm] {\textoverline{{\tt onCancelled()}}} (4) ;
    \path[->] (1) edge[loop right, looseness=5] node[right] {{\tt cancel()}          } (1) ;
    \path[->] (3) edge[loop below, looseness=10] node[below] {{\tt cancel()}          } (3) ;
    \draw[->,rounded corners]
            (2)
            -- (0 |- 0,-2.6cm)
            -- (1 |- 0,-2.6cm)
            node[below] {\textoverline{{\tt onPostExecute()}}}
            -- (4 |- 0,-2.6cm)
            -- (4) ;
  \end{tikzpicture}
  \caption{Learned typestate of the AsyncTask class}
  \label{fig:AsyncTask}
\end{figure}
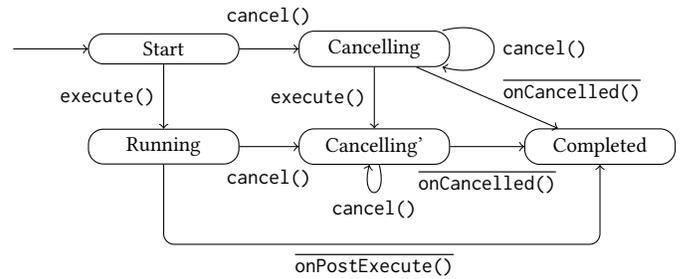

Our experiment revealed an unexpected edge-case: if \callin {execute}
is after \callin {cancel} but before the \callback{onCancelled}
callback is received, it will not throw an exception but will never
cause the callback task to be run.  The learned interface is
in Figure~\ref{fig:AsyncTask}.

\experhd{SpeechRecognizer} 
This class provides an example of uncontrollable environmental
non-determinism.  The particular callback that signals the end of the
speech session---either an \callback {onResults} or an \callback
{onError}---is determined by the environment (in particular, the sound
around the phone during the test).  In this case, to reduce the system
to a deterministic one we can learn, we supposed that the state after
an \callback {onResults} or \callback {onError} is the same and merged
the two callbacks into a single \callback {onFinished} symbol.

Our results revealed two interesting corner cases for the ordering
of inputs.  First, if an app calls \callin {cancel} between
calling \callin {startListening} and receiving the \callback
{onReadyForSpeech} callback (represented by our ``starting''
output symbol), calling \callin {startListening} again will have
no effect until after a certain amount of time, as shown by the
{\tt wait} transition from state ``Cancelling'' to ``Finished''.
Delays in readiness like this can be generally considered bugs; if a
system will not be ready immediately for inputs it should provide a
callback to announce when the preparations are complete, so as not
to invite race conditions.

Our second corner case is where the app calls \callin
{stopListening} as the very first input on a fresh
SpeechRecognizer.  This will not throw an exception, but calling
\callin {startListening} at any point after will fail, making the
object effectively dead.

\experhd{SQLiteOpenHelper} This class provides a more structured
interface for apps to open and set up SQLite databases.  It has
callbacks for different stages of database initialization,
allowing apps to perform setup operations only as they are needed.
When a database is opened with \callin {getWritableDatabase}, a
callback \callback {onConfigure} is called, followed by an
\callback {onCreate} if the database didn't exist yet or an
\callback {onUpgrade} if the database had a lower version number
than was passed to the SQLiteOpenHelper constructor, all followed
finally by an \callback {onOpen} when the database is ready for
reading.  The database can then be closed with a \callin {close}.

Our experiment observed the callbacks when opening
databases in different states (normal, non-existent, and out of
date) and performing the \texttt{close()} operation at different
points in the sequence.  We found that once the
\texttt{getWritableDatabase()} method is called, calling
\texttt{close()} will not prevent the callbacks from being run.

\experhd{VelocityTracker}
This class was a special case with no asynchronous behavior; it was a
test of our tool's ability to infer traditional, synchronous
typestates.
The class has a \callin{recycle} method that we expected to disable
the rest of the interface, but our tool found (and manual tests
confirmed) that the other methods can still be called after recycling.
The documentation's warning that ``You must not touch the object after
calling [recycle]'' is thus not enforced.

\section{Related Work}
\label{sec:relwork}

Works which automatically synthesize specifications of the valid
sequences of method calls
(e.g.~\cite{ACMN05,shoham-yahav,popl-spec-mining,dynamic-alternating-rules})
typically ignore the asynchronous callbacks.

Static analysis has been successfully used to infer typestates
specifications (importantly, without callbacks)~\cite{ACMN05,henzinger-synthesis,shoham-yahav}.
The work in~\cite{ACMN05} infers classical typestates for Java classes
using \LStar.
In contrast, our approach is based on testing.  Therefore, we
avoid the practical problem of abstracting the framework code. On the
other hand, the use of testing makes our \LStar oracles
sound only under assumptions.
Similarly, ~\cite{Giannakopoulou2012} uses \LStar to infer classical
typestates, including ranges of input parameters that affect
behavior. However, their tool is based on symbolic execution,  and thus
would not scale to systems as large and complex as the Android
Framework.

Inferring interfaces using execution traces of client programs using
the framework is another common
approach~\cite{popl-spec-mining,whaley-spec-mining,acharya-fsm-mining,dallmeier-fsm-mining,gabel-fsm-mining,walkinshaw-ltl-fms-mining,perracotta,DBLP:conf/kbse/PradelG09}.
In contrast to dynamic mining, we do not rely on the
availability of client applications or a set of execution traces.
The \LStar algorithm drives the testing.

The analysis of event-driven programming frameworks has recently
gained a lot of attention (e.g.
~\cite{DBLP:conf/pldi/ArztRFBBKTOM14,DBLP:conf/oopsla/BlackshearCS15,DBLP:conf/ndss/CaoFBEKVC15,node-js-cg}).
However, none of the existing works provide an automatic approach to synthesize
interface specifications.
Analyses of Android applications mostly focus on either statically proving
program correctness or security
properties~\cite{DBLP:conf/pldi/ArztRFBBKTOM14,DBLP:conf/oopsla/BlackshearCS15,DBLP:conf/ndss/GordonKPGNR,DBLP:conf/ccs/WeiROR14,DBLP:conf/sigsoft/FengADA14}
or dynamically detecting race
conditions~\cite{droidracer,cafa,DBLP:conf/oopsla/BielikRV15}.  These
approaches manually hard-code the behavior of the framework to increase the precision of the analysis.
The callback typestate specifications that we
synthesize can be used here, avoiding the manual
specification process.

Our work builds on the seminal paper of Angluin~\cite{Angluin87} and
the subsequent extensions and optimizations. In particular, we build on
\LStar for  I/O automata~\cite{AV10,ShahbazG09}. The
optimizations we use include the counterexample suffix analysis
from~\cite{RivestS93} and the optimizations for prefix-closed
languages from~\cite{HNS03}.
The relation to conformance testing methods~\cite{Wmethod,WPmethod,Hennie64,Gonenc70,SabnaniD85} has been discussed in Section~\ref{sec:equiv}.

\vspace{-1ex}
\section{Conclusion}
\label{sec:conclusion}
We have shown how to use active learning to infer callback typestates. We
introduce the notion of distinguisher bound which take advantage of good
software engineering practices to make active learning tractable on the Android
system. Our method is implemented in the freely available tool called \tool.
This paper enables several new research directions. We plan to
investigate mining parameters of callins from instrumented trace from real user interactions, as well as the inference of structured typestates (for instance, learning a typestate as a product of simpler typestates).

\vspace{-1ex}
\begin{acks}
This research was supported in part by DARPA under agreement
FA8750-14-2-0263.
Damien Zufferey was supported in part by the European Research Council Grant
Agreement No. 610150 (ERC Synergy Grant ImPACT
(\url{http://www.impact-erc.eu/})).
\end{acks}

\bibliographystyle{plain}
\bibliography{main}

\end{document}